\documentclass[preprint,authoryear,12pt]{elsarticle}
\usepackage{amssymb}

\journal{New Astronomy}

\begin{document}

\begin{frontmatter}

\title{Effect of Blue Component Stars in parameters of NGC 6866}

\author{Gireesh C. Joshi}
\ead{gchandra.2012@rediffmail.com}

\address{P.P.S.V.M.I. College, Nanakmatta, U.S. Nagar-262311, Uttarakhand, India}

\begin{abstract}
The open star clusters (OSC) are important tracers for
understanding the Galactic evolution. The parametric study
of these astronomical-objects are crucial task due to the
appearing sequence of the members of OSC. These members
are defined through the various approaches such as
photometric, statistical, kinematics etc. In the present paper,
we have been using the photometric colours of the identified
stars for categorized them into the blue and red component
groups and identification of these groups is possible through
(B-V) vs V colour magnitude diagram (CMD). Furthermore,
the influence/effect of these groups is also examined in the
estimation of cluster parameters. The stellar enhancement of
cluster NGC 6866 is found through the blue-component-stars
(BCS) and the linear solution of best fitted values of King
Models of the radial-density-profile (RDP) gives the core
radius as
 $5.22{\pm}0.29~arcmin$. The good agreement of present estimated parameters of the cluster with the literature  seems to be an effective evidence to consider BCS as the true representative of the cluster. The stellar distribution of the cluster shows continuous phenomena of the mass segregation. An effect of incompleteness of photometric data is well understood in the term of the mass-function slope values, which is found to be $-3.80{\pm}0.11$ and also shows increment nature with the incompleteness. 
\end{abstract}

\begin{keyword}
(Galaxy): Open star cluster; individual: NGC 6866, blue-component; Most probable members
\end{keyword}

\end{frontmatter}


\section{Introduction}
\label{}
The observational study of intermediate-age open clusters plays an important role to constrain the theories of stellar and Galactic evolution \citep{sha06}. Though, the cluster region is also influenced by field stars, therefore, the identification of probable members of open-star-clusters (OCs) becomes a crucial task for parametric analysis. The influence of field stars within the cluster may be occurred due to embedded position of the cluster on the Galactic disk, which easily seems through the stellar distribution on colour-magnitude diagram (CMD) and the proper motion plane. Such influence of field stars known as contamination, which depends on the location of a particular cluster on Galactic plane \citep{jos14}. As a result, scattered stellar distribution of stars may be occurred in the CMD and definitely produced the photometric broadening of members in the sequence of stellar evolution. Such broadening in appeared stellar sequence of the CMD may be produce the uncertainty  in exact identification of the stellar sequence, which leads to estimation error in the fundamental parameters (reddening, distance and age) of the cluster. It is believed that the accurate measurement of all above said parameters will be possible after decontamination of the field stars from cluster region. The above said decontamination would be carried out through either the photometric statistical criteria \citep{sha08} or the statistical cleaning in CMDs \citep{jos14}. Recently, \citet{jos15} have been introduced the statistical cleaning procedure of field star decontamination through the colour-magnitude distance for field star decontamination, which is dependent on the grid size around the field star. The size of grid would be fixed on the basis of comparison of cluster and field stars in the term of  their magnitude and colour \citep{jos15}. After field stars decontamination from the observed CMDs, it becomes very effective tool to derive the precise values of fundamental parameters with the comprehensive  results of the Luminosity-function (LF) and the Mass-Function (MF).\\
The outlines of this paper is described as following. The quality of available photometric data of cluster has been prescribed in Section~\ref{s:qua}. The separation procedure of blue and red component stars of this cluster is described in Section~\ref{s:blue_red}. The RDPs and SNDPs of both type stars have been described in Section~\ref{s:rdp_blue} and Section~\ref{s:snd}, respectively. The identification procedure  of giant star sequence within cluster have been discussed in Section~\ref{s:sdm}. The estimation of reddening, distance and age are described in Sections~\ref{s:pm} and \ref{s:cmd}. The MF study and examination of nature of total-to-selective-extinction are done in Sections \ref{s:mf} and \ref{s:tcd2} respectively. Finally, we have been summarized our results and discussed their importance in Section~\ref{s:con}.
\section{The previous studies and data}\label{s:qua}
Recently, three photometric catalogue of deep CCD observations are available for this cluster. These catalogues are constructed by \citet{jos12}, \citet{jan14} and \citet{bos15}. The field of view of these observations are about 13.5$\times$13.5 arcmin$^2$, 42.8$\times$42.8 arcmin$^2$ and 21.5$\times$21.5 arcmin$^2$, respectively. All authors of these catalogue have been derived the fundamental parameters of cluster based on their adopted algorithm procedure. \citet{bos15} have been shown comparison of their photometry with the \citet{jos12} and \citet{jan14} in their manuscript at Figure 4 and Figure 5 respectively. On the basis of above prescribed figures, it is concluded that their photometry and \citet{jan14} are well matched to each other. Since, \citet{jan14} is contained the more field of view of the cluster region rather than others, therefore, it is used to further analysis of influence of blue-component-stars (BCS) in the estimation of cluster parameters. The first part of present manuscript represents the role of BCS to constrain the dynamics of cluster.\\ 
\begin{figure}
\includegraphics[width=13.5cm]{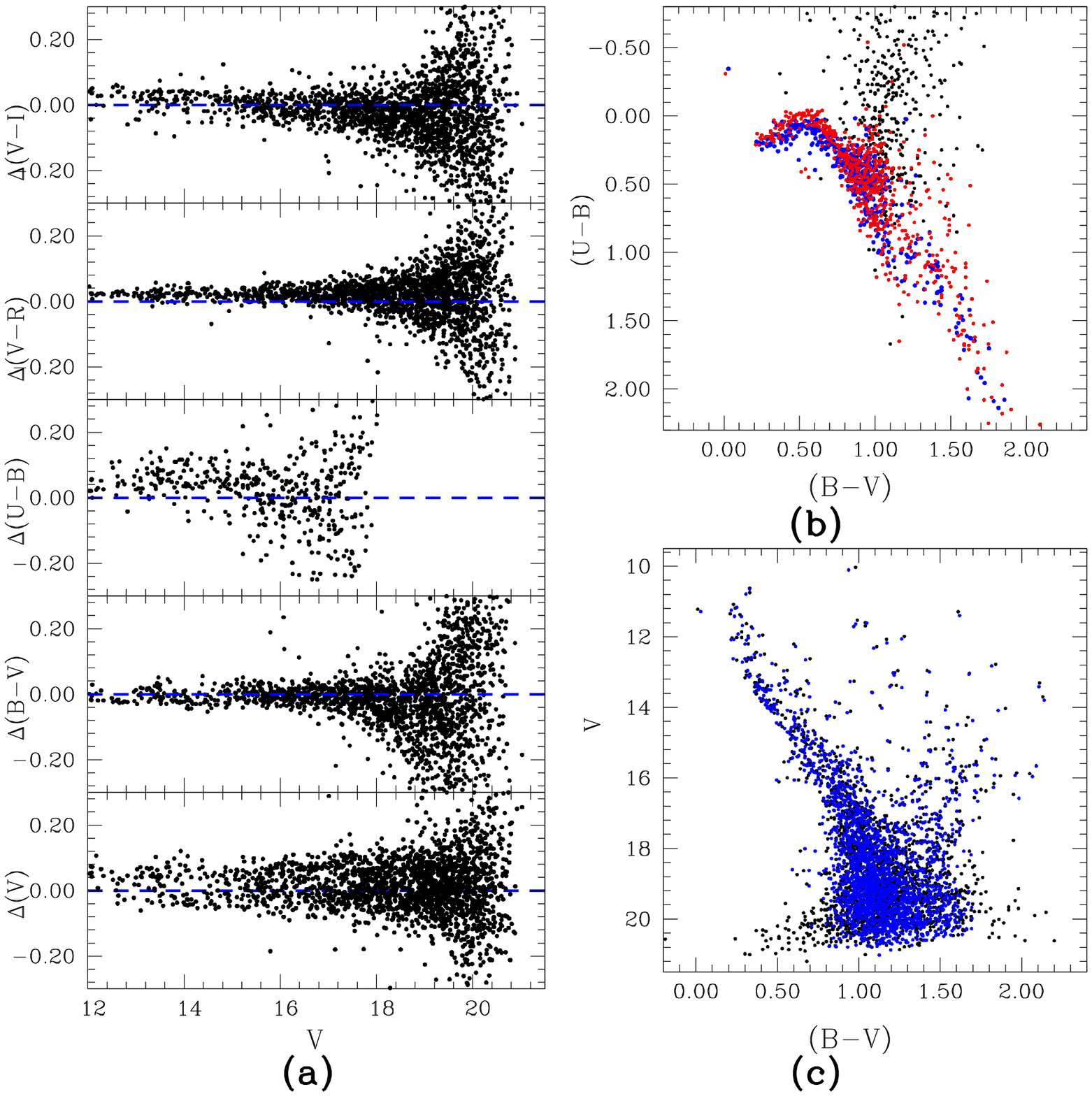}
\caption{{\bf a)} Comparison of photometry catalogues between \citet{jos12} and \citet{jan14}. {\bf b)} The comparison of $(U-B)$ vs $(B-V)$ two-colour diagrams of these catalogues. {\bf c)} Here, We have also shown $(B-V)$ vs $V$ CMDs from both cataloges. The black and blue dots of panels (b) and (c) are represent stars from \citet{jos12} catalogue and \citet{jan14} catalogue respectively. The red dots of panel (b) are those stars of \citet{jos12} catalogue, which having $V$-magnitude less than 18 mag.}
\label{s:fig01}
\end{figure}
To see the difference of stellar magnitude of common stars between \citet{jan14} and \citet{jos12}, we are cross-matched the photometric magnitude of stars within scatter of $1~arcsec$. After comparison, we have found 2092 common stars between both catalogues. Furthermore, we have been obtained a slope in residual magnitudes of both catalogues, which has been shown in the residual magnitude vs $V-$magnitudes plots (as depicted in the left panels of Figure 1). A deep investigation of figures 4 and 5 of manuscript of \citet{bos15} has clearly shown that their photometry shows approximate zero residual with \citet{jan14} and non zero magnitude residual with \citet{jos12}. On the behalf of these figures, it seems to be some technical problems with the procedure of estimation of stellar magnitude of the work of \citet{jos12}. For further investigation for it, we have been constructed the plot of $(B-V)$  vs $(U-B)$ TCD (as shown in Figure 1-b). The stellar spread is clearly observed in this TCD, which has been produced by those stars which having $V-$magnitude greater than 18 mag. In this figure, we have also seen a upward shift of the stellar distribution (depicted by red dots) through \citet{jos12} catalogue in the comparison of catalogue of \citet{jan14}. This shifted colour values of stars increase the reddening value of the cluster, which is easily noticed by the comparison of resultant values of reddening from the both works.\\
In the Figure 1-c, we have been depicted the stellar distribution on the CMD through the magnitude extracted from the both catalogue. The black and blue dots of above described figure represents the stars from \citet{jos12} and \citet{jan14} respectively. It is seems to be approximate similar stellar distribution in CMD except lower end of CMD, which leads to the slightly change in the value of apparent distance modulus.The value of reddening, $E(B-V)$, is required for calculation of cluster distance. Reddening value of cluster is obtained by fitting Zero-Age-Main-Sequence (ZAMS, \cite{sck1982}) in the $(U-B)$ vs $(B-V)$ diagram as shown in Figure~\ref{s:fig07}. In this figure, the black curve shows the reddened ZAMS which is obtained after the adding a shift 0.12 mag in $E(B-V)$ values of normal ZAMS.
\begin{figure}
\includegraphics[width=13.5cm]{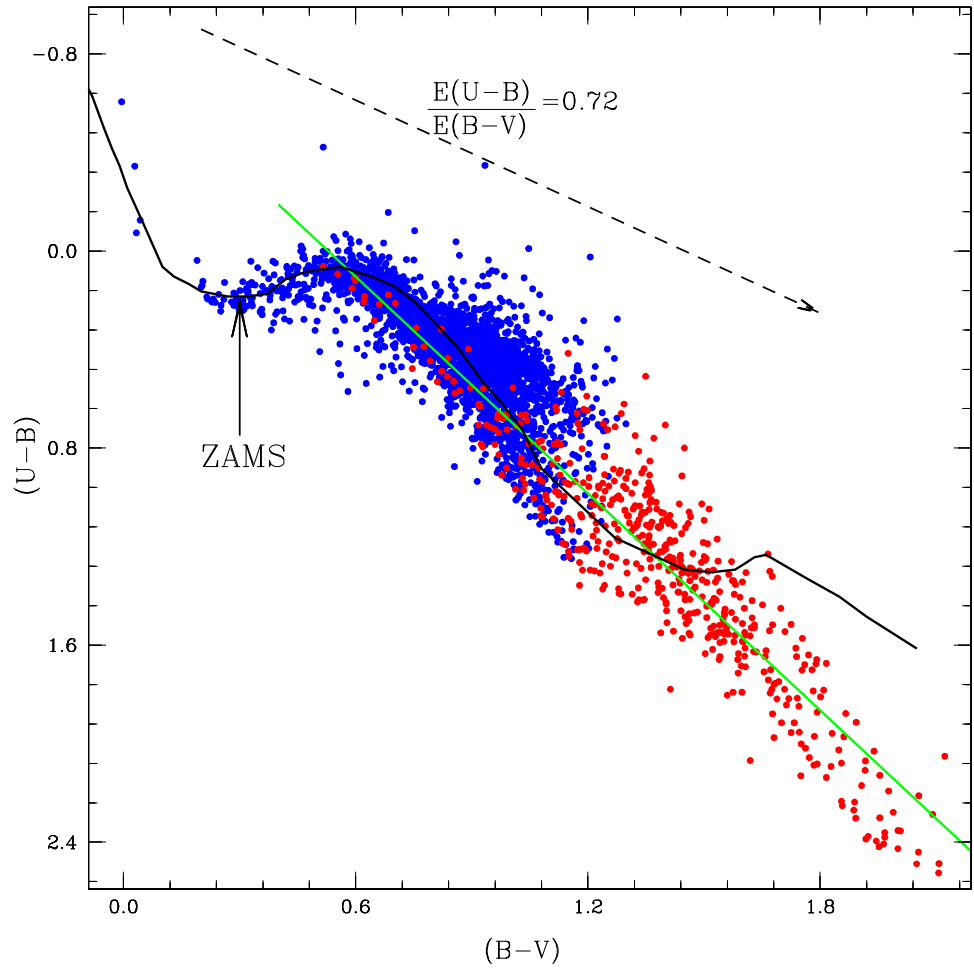}
\caption{The blue and red component stars of cluster is depicted by blue and red dots respectively. The solid black and green line represent the best fitted ZAMS on blue component stars and best linear fit in red component stars of cluster. The dashed arrow shows normal reddening law.}
\label{s:fig07}
\end{figure}
The reddened ZAMS fitting is occurred for BCS while the RCS are show a linear trend. We have fitted a linear equation to obtained linear relationship between two colours of RCS.
The solution of this linear equation is found to be as,
$$(U-B)=(1.47{\pm}0.03)(B-V)-0.78{\pm}0.04.$$
Though, the RCS stars, having ($B-V{\geq}1.8~mag$), found to be below this linear solution. As a result, the slope of $(U-B)/(B-V)$ is found to be very high for RCS compare to expected slope for the clusters.\\
It is noticeable that the scattering for $A_0$ type stars found to be less and increases with ($B-V$) after $A_0$ type stars. The arrow indicates the normal colour excess law ($\frac{(U-B)}{B-V}=0.72$, which is used for adding a shift in colour-excess $(U-B)$. Since, some RCS are also giant members of cluster, therefore, the shifting value chosen in such a way that the redden ZAMS also follow the pattern of RCS within the two colour region of BCS. Some RCS are embedded in the distribution of BCS on the TCD. Mostly such type RCS are shows below position from best fitted ZAMS on the TCD and they have lesser scatter.  then they give the opportunity to find best fitted reddened ZAMS together with $A_0$ type stars of the cluster. Thus, the reddening of cluster is found to be $0.12{\pm}0.01$ mag which is slightly high compare to reddening (0.10 mag) estimated by \cite{jos12} and shows close agreement to reddening ($0.16{\pm}0.04$)as derived by \cite{jan14} through Bayesian analysis.  
\section{Separation procedure of Blue and Red component stars}\label{s:blue_red}
\begin{figure}
\includegraphics[width=13.5cm]{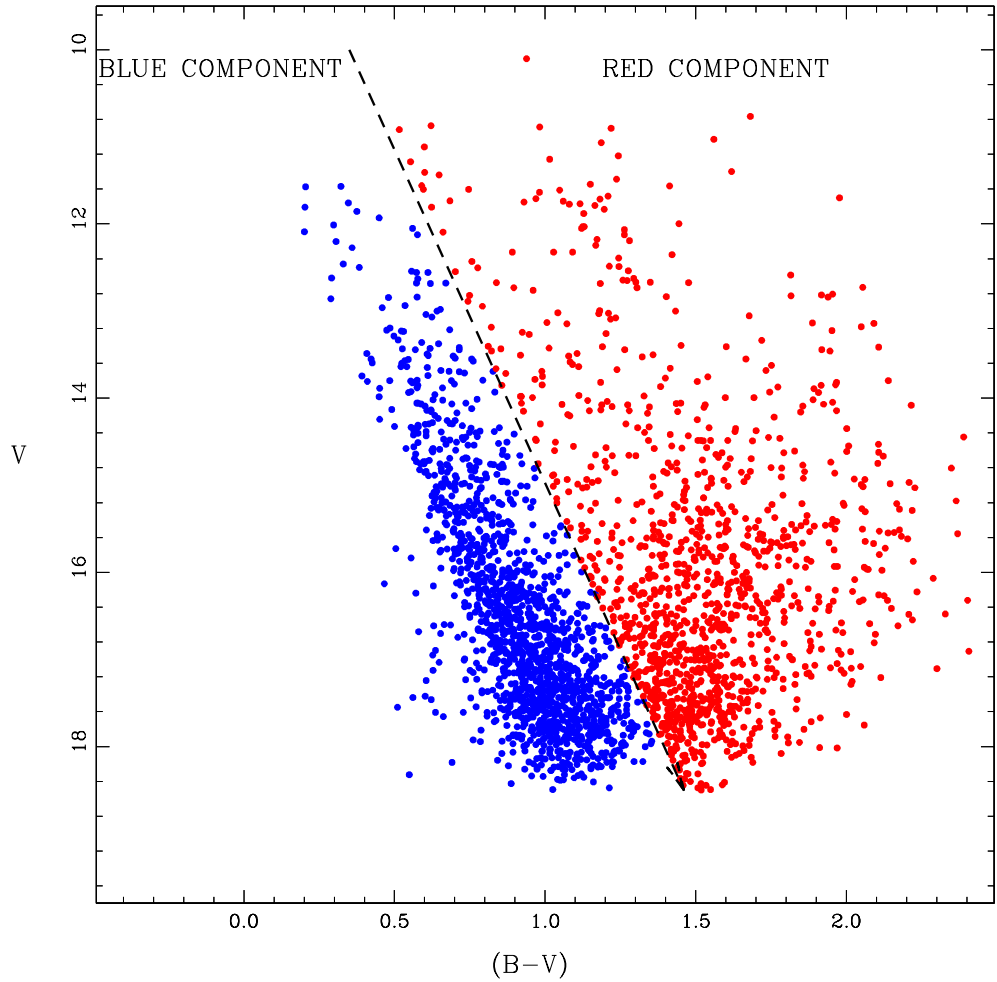}
\caption{The blue and red component stars of the cluster have been shown by blue and red dots. The black dashed arrow represents the separation boundary of these two type components.}
\label{s:fig02}
\end{figure}

The stellar distribution seems to be spread out on (B-V) vs V CMD (as shown in Figure~\ref{s:fig02} and Figure~\ref{s:fig11}). The dense stellar distribution is obtained to be broad and sloppy line as shown in the left side of the above prescribed CMD, which may be occurred due to the evolutionary processes of the probable main sequence (MS) of the cluster. The comparison of known literature characteristic with the cluster characteristics through this sequence (detailed discussion in at Section~4) is verified the fact that said sequence is the MS of present studied cluster. Other hand, such linear trend does not appear for the stellar distribution of right side of CMD but the space is sparsely fulfilled by other detected stars of the cluster. The said sloppy sequence of the cluster is clearly identified through visually inspection of the distribution of massive and fainter stars. Here, We are used following magnitude-colour relation to separate this probable MS from the stellar distribution of the observed field of cluster,
$$V_{o}=7.66{\times}(B-V)_{o}+7.32,$$
where $V_{o}$ and $ (B-V)_{o}$ are the observed magnitude and colour of stars.      
The above prescribed linear equation is represented by a black dashed arrow on $(B-V)$ vs $V$ CMD of the cluster as depicted in Figure~\ref{s:fig02}.
The left sided stellar distribution of this arrow is referred as the distribution of blue component stars (BCS), whereas stars of right side of this arrow is considered to be red component stars (RCS) of the cluster. The blue and red dots in Figure~\ref{s:fig02} are represented the BCS and RCS of the cluster. The obtained RCS may be either super-giant stars of the cluster or field stars. The super-giant stars are those cluster members which are evaporated from the main sequence through the dynamical evolution phenomena of the cluster. A total of 4088 stars and 1250 stars are found to be blue component and red component stars, respectively. This separation procedure is applied for those stars which are brighter than 18.5 mag in V-band due to the following fact.
The CMDs in Figure~\ref{s:fig11} is direct indicate that BCS and RCS are not easily separated to each other for those stars, which having V-magnitude greater than 18.5 mag. Although it seems that RCS of $(V-I)/V$ CMD are visibly separated from the MS stars but are not seen separately in the $(B-V)/V$ CMD.

\section{Cluster extent due to blue component Stars}\label{s:rdp_blue}
\begin{figure}
\includegraphics[width=13.5cm]{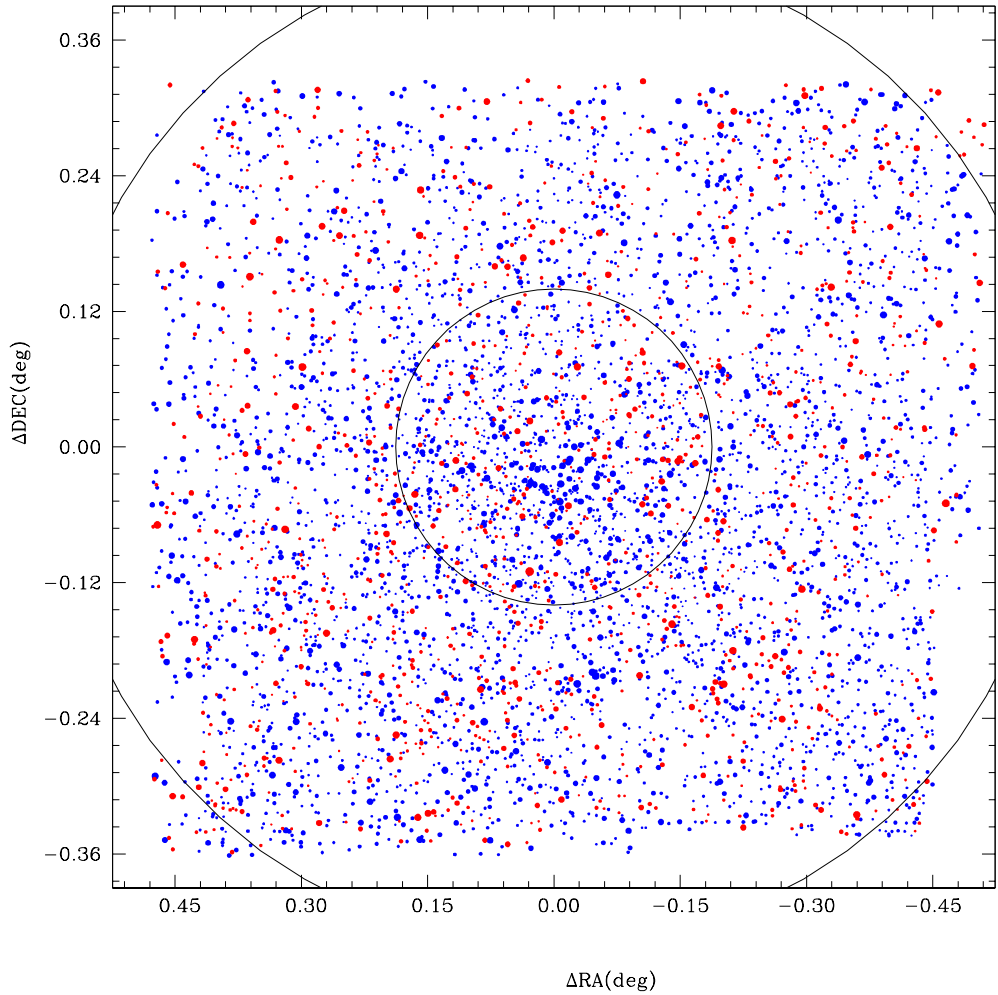}
\caption{The inner and outer circles represent the core radius and limiting radius of clusters. The size of the dots is proportional to their brightness. The blue and red dots are shown to be blue and red component stars respectively.}
\label{s:fig03}
\end{figure}

\begin{figure}
\includegraphics[width=13.5cm]{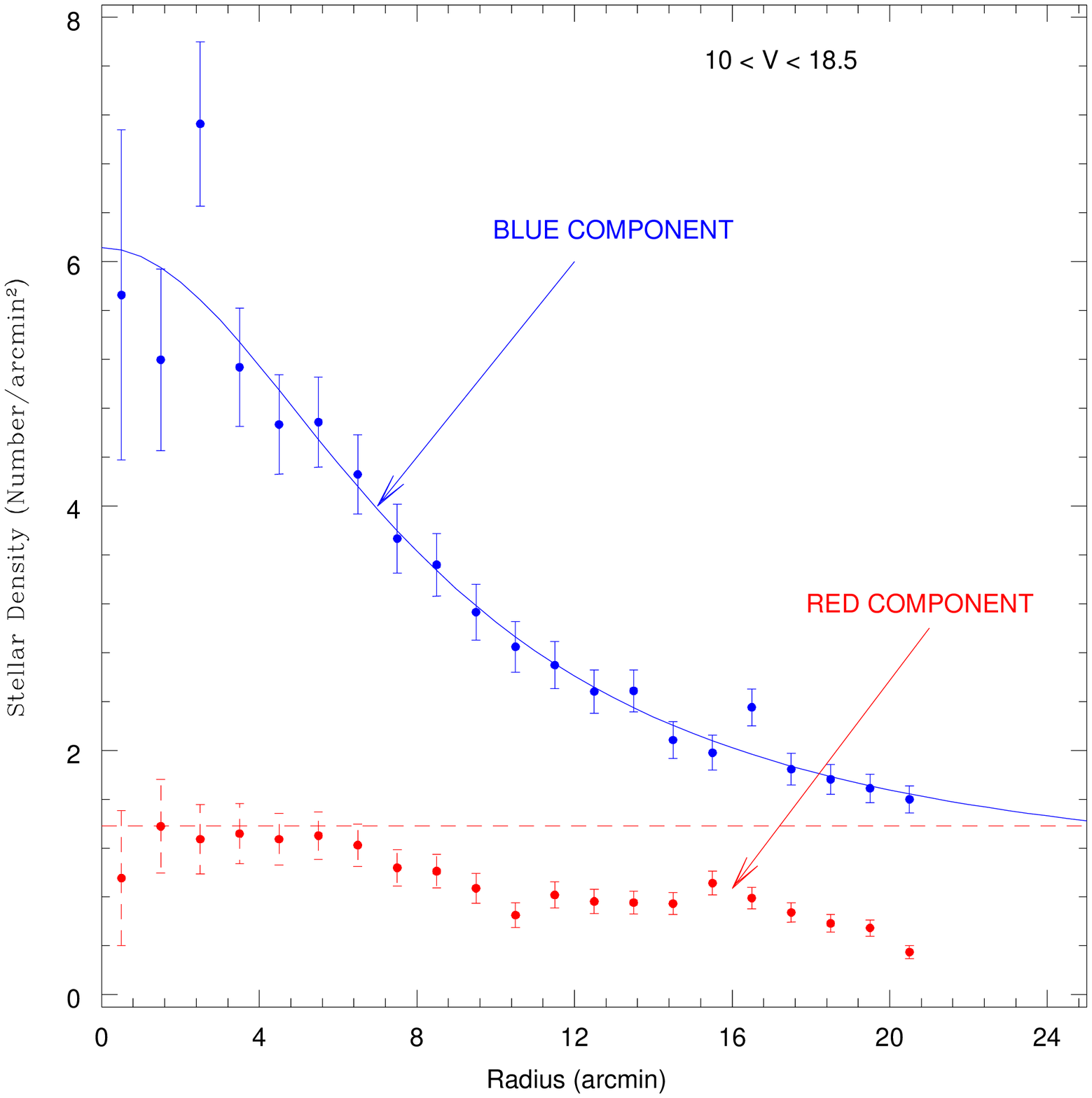}
\caption{The blue and red component stellar density of the cluster has been shown by blue and red dots respectively.}
\label{s:fig04}
\end{figure}

The finding chart is very effective to check out that whether the BCS and RCS are present on a particular sky distribution of cluster region or not. The stellar distribution of BCS and RCS are shown in Figure~\ref{s:fig03}. In this figure, the size of dots is the function of the stellar brightness and increases towards to the lower magnitudes of the stars. The smallest size of dots is represented those stars which having V-magnitude to be 18.5 mag. Our deep analysis shows that the model limiting radius is greater than that of the observed cluster field by \textsf{\cite{jan14}}. The procedure of estimation of limiting radius is prescribed as below.\\ 
The cluster center is taken in such way that we have obtained best RDP through it and follow the condition of maximum count within the cluster region. The detail procedure of above said approach is prescribed by the \citet{jos15}. In the present case, the cluster center $(\alpha,~\delta)$ is found to be ($20^h:03^m:54.9^s,~+44^o:09^{'}:28.2^{''}$) in $RA-DEC$ plane. This coordinate is  slightly shifted from the value of cluster center as listed at SIMBAD database and various concentric rings are constructed around the center to determine the radius of the cluster. The radial densities of both components are computed for each ring which are used to construct the radial density profiles  for both components as shown in Figure~\ref{s:fig04}.
The above prescribed profile of stars of both component is shown direct evidence of cluster characteristics through BCS. The following empirical formula of King Model \citep{kin62} is fitted on the RDP through BCS of cluster, $$\rho(r)=\rho(0)+\frac{f_0}{1+(\frac{r}{r_c})^2},$$ 
where $f_0$, $r_c$, $\rho(0)$ and $\rho(r)$ are peak stellar density, core-radius, background stellar density and instantaneous stellar density, respectively. The value of $f_0$, $r_c$ and $\rho(0)$ for the BCS are found to be $5.21{\pm}0.29$, $8.39{\pm}0.85$ and $0.89{\pm}0.16$, respectively. The resultant fit cuts the $\rho(0)$ at the outside of range of determined radial densities, therefore, we are not capable to give the value of cluster radius. The exact cluster radius would be defined after the complete observation of the field view of the cluster. Other hand, the value of limiting radius [$r_{limiting}=r_c \sqrt{\frac{f_0}{3 \sigma}-1}, ~\sigma$ is the estimation error in background stellar density] can be obtained without any cut point. Hence, we obtained the limiting radius of the cluster which comes as $26.48{\pm}0.32~arcmin$. The values of radial densities of RCS did not find in the order of continue decreasing as seems for the BCS. The RDP of RCS seem to be made by two plateau regions, such as, $0<r{\leq}7~arcmin$ and  $10<r{\leq}17~arcmin$. The visual inspection of stellar densities of these regions of RCS shows that the first one (core) is higher stellar density than that of the later one (outside). The higher stellar density of RCS at the core region may be appears due to the enhancement of probable super giant members of the cluster. Furthermore, the sloppy region of RCS is found to be at the radial distance range of $7<r{\leq}10~arcmin$, which may be the core-corona transition region of the cluster. Similarly, the model core radius limits, $7.54<r_c<9.24$, may also be clue for the existence of such type region.   
\section{Surface number density tracer of mass segregation}\label{s:snd} 
Surface number density profile (SNDP) is very effective to shown the stellar variation of various magnitude bin with the radial distance. Such information will be becomes an ideal tracers to construct the evolutionary models of stellar evolution, which is beneficial to understand the stellar dynamics of the cluster. In addition, SNDPs of various radial zones are effectively explain the possible stellar distribution according to their masses. Similarly, this profile also helpful to understand the over-density inside the core-radius of cluster and possible cause of constant stellar density of RCS after the core region of cluster. The SNDPs of both components have been depicted in Figure~\ref{s:fig04}.\\
\begin{figure}
\includegraphics[width=13.5cm]{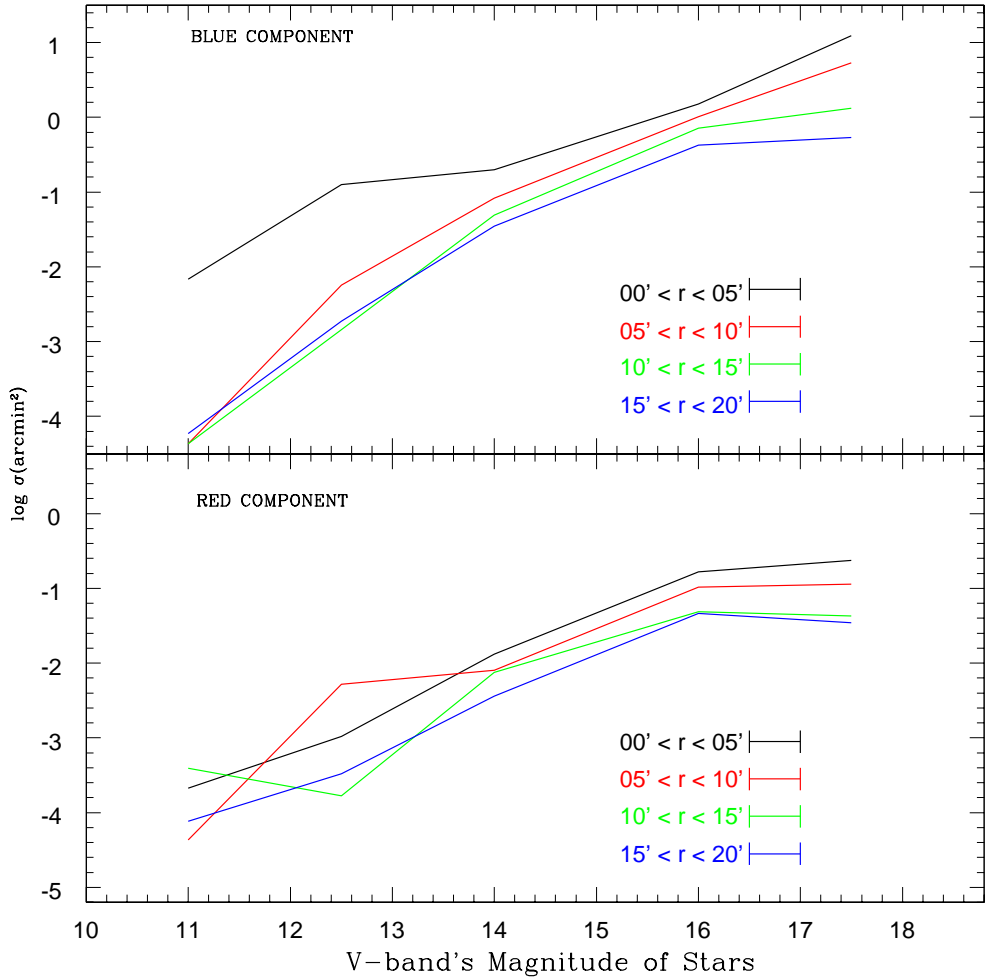}
\caption{The upper and lower panels are shown SNDP for BCS and RCS respectively. The surface number density is defined as the number of stars per magnitude interval in unit area \textsf{\citep{sun96}}.}
\label{s:fig05}
\end{figure}
The SND values of the BCS indicates that mostly massive stars are concentrated at the innermost region ($0<r{\leq}5~arcmin$) of the cluster in the comparison of the fainter stars. The adjacent SNDPs of BCS of two innermost ring regions shows clear separation between them and this separation continuously  decreases upto 16 mag of $V-$band whereas it increases for those members which having stellar magnitude greater than above prescribed magnitude limit. This result shows the evidence of mass-segregation phenomena within cluster. The SNDPs of stars (having stellar $V-$magnitude less than $14~$mag) of outermost region indicate that the  SND values of massive stars do not distinguish from each other. For other stars, these profiles are distinguishable and their SND values decrease in the order of number of rings from the cluster center.\\
The SNDPs of RCS of core region shows clear separation between them and both profiles is stellar enhancement compare to the SNDPs of coronal region of the cluster. The profiles of coronal region are not so much visually separated as seen for former profiles. It is also noticeable fact that the values of SND of massive stars are obtained to be close to each other for both category (BCS and RCS). Such stellar distribution gives an indication of similar birthing environment of massive stars of both components. Due to this fact, it may be concluded that some massive stars are evaporated from the MS evolution of the cluster but they are still belong to the cluster. These so called stars may be the super-giant members. It can be understandably fact that the density of massive stars of field region too high as for massive members of the cluster. The higher density of massive members is possible due to the addition of super-giant members of the cluster with field stars. The slightly high value of radial density of RCS of the core with compare to the densities than that for corona (as depicted in Figure 4), may be confirm the mixing of super-giant members of the cluster with the field stars. On the behalf of this discussion, it may be concluded that the all RCS of the coronal region are field stars due to the absence of massive stars. 

\section{Identification of cluster giant stars through statistical distance method}\label{s:sdm}
Since, the RCS of core region of clusters contains the super-giant stars with filed stars whereas the RCS are coronal region are only field stars. We are assumed that the field stars of RCS are uniformly distributed over all the cluster region. To identify the possible giant stars of the cluster, we have been applied the statistical distance algorithm on the RCS [detailed procedure has been prescribed by \cite{jos15}]. The core ($0<r{\leq}8.39~arcmin$) is considered to be giant stars region and the region, having radial distance range ($10<r{\leq}13.05~arcmin$) and equivalent area to giant star region, considered as the filed region. Here,  we are used $V$, $B$ and $I$ filters instead of $J$, $H$ and $K$ filters and the magnitude-colour distance is taken as,
$$D_{mc}= \sqrt {\Delta_{V}^2+ \Delta_{BV}^2 + \Delta_{VI}^2},$$  
where $\Delta_{V}$, $\Delta_{BV}$ and $\Delta_{VI}$ are the differences of $V$ magnitude , $B-V$ colour and $V-I$, respectively, between stars of field region and giant (core of cluster) region. For it, the $\Delta_{V}$, $\Delta_{BV}$ and $\Delta_{VI}$ have been considered to be $0.50$, $0.15$ and $0.15$ respectively. After applying this procedure, we have found 129 giant stars among 264 RCS of the core region. The giant star vertical sequence presents at the colour, ($V-K$)= 4.0 mag of Figure~\ref{s:fig08}. In this Figure, a parallel giant star sequence of main sequence stars is also being appeared. These both giant sequences are depicted by red dots. The position of these sequence stars is indicated that the giant stars are born in a different time span and different dynamical evolution processes.\\
It is also noticeable fact that all BCS members are not main sequence of stars and some of them may be field stars. These stars are not possible to separate from MS due to following facts. Firstly, the available photometric data not cover all cluster region. Secondary, photometrc criteria \citep{sha08} does not tell about the field stars, which having colour and magnitude values in the similar order of the MS of cluster.
\section{No clue for distinguish of stars through Proper Motion}\label{s:pm}
The members of the cluster are loosely gravitational bound and each members shows its individual motion, therefore, members change their location in sky with respect to each other. The exact determination of velocity of stars are not possible through observations but parallax used to determine the proper motion (angular displacement) of stars by using the datasets of two different time span. Here, the proper motion values extracted from the catalogue of \citet{ros10}. This catalogue provides position in the International Celestial Reference System (accuracy: 80-300 mas) and absolute proper motion (accuracy: 4-10 mas yr$^{-1}$) of about 900 million stars, derived from the optical USNOB1.0 and near-infrared 2MASS catalogues. We have been estimated the mean proper motion of BCS and RCS as ($-2.02{\pm}0.15~mas~yr^{-1}$, $-4.52{\pm}0.16~mas~yr^{-1}$) and ($-2.18{\pm}0.25~mas~yr^{-1}$, $-4.24{\pm}0.27~mas~yr^{-1}$), respectively. The mean proper motion values of both components have been computed through utilizing the proper motion value of 973 stars among 1093 BCS and 291 stars among 342 RCS, respectively. Their vector-point diagrams are shown in Figure~\ref{s:fig05}.
\begin{figure}
\includegraphics[width=13.5cm]{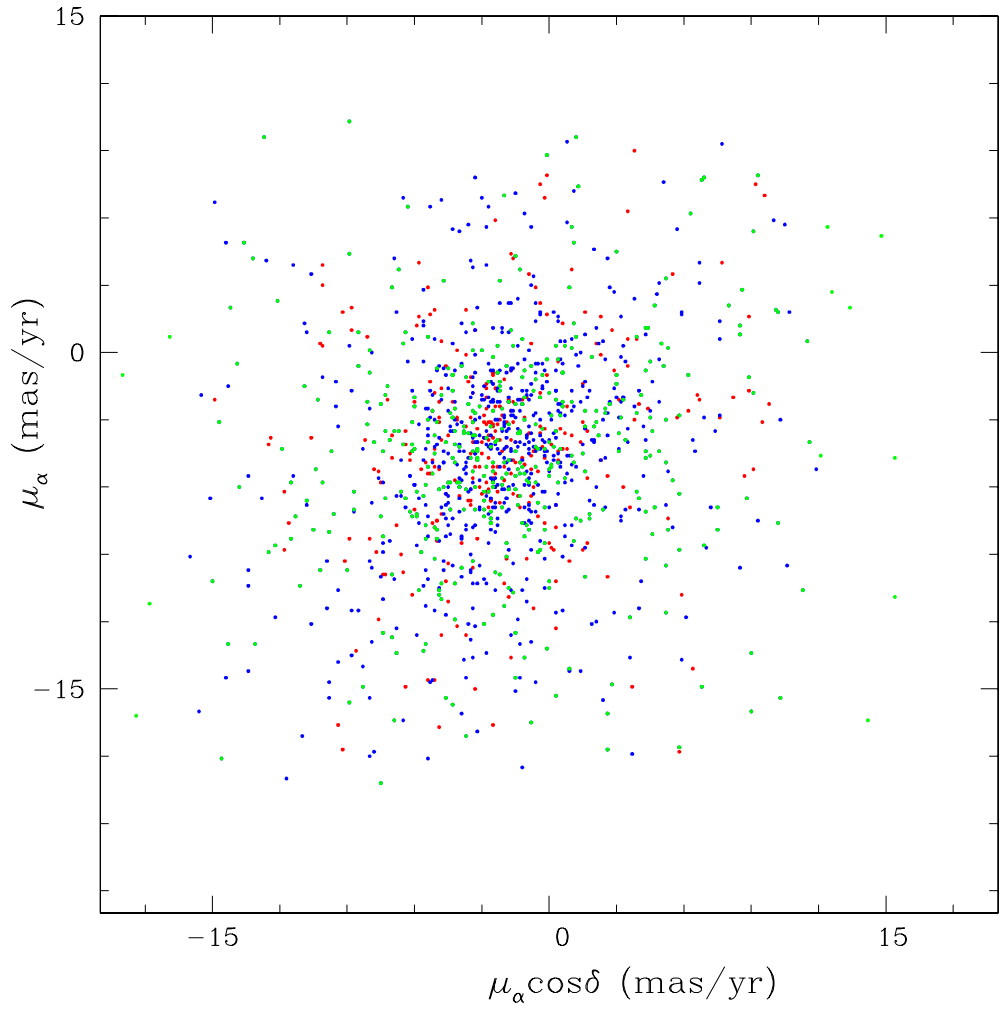}
\caption{The proper motion distribution of BCS and RCS is depicted by blue and red dots respectively.}
\label{s:fig06}
\end{figure}
We have not found any clear trend separating members and field stars, such type result is also found by \citet{bel10} through the study of nearby well-studied open star clusters. The mean proper motion of RCS has slightly differ from that of BCS which happened due to influence of field stars in RCS and we have already shown in Section~\ref{s:sdm} that RCS  of core region are mixture of the cluster giant and field stars.\\     
Since, it is well known fact that the influence of field stars is high in coronal region compared to core. Due to the high membership probability, we have also estimated the mean proper motion of the BCS of core using proper motion values of 411 stars which are left over after applying an iteration procedure (as suggested by \cite{jos14}) on 458 blue component members of the core-region. These value are coming to be $-1.78{\pm}0.28~mas~yr^{-1}$ and $-4.58{\pm}0.25 mas~yr^{-1}$ in Right Ascension (RA) and Declination (DEC) directions, respectively, and members are depicted by green dots in Figure~\ref{s:fig06}. The mean proper motion values of core and whole cluster is found to be similar within uncertainty which suggested that the BCS of the cluster are less influenced by field stars.

\section{Distance and Age of the cluster}\label{s:cmd}
\begin{figure}
\includegraphics[width=13.5cm]{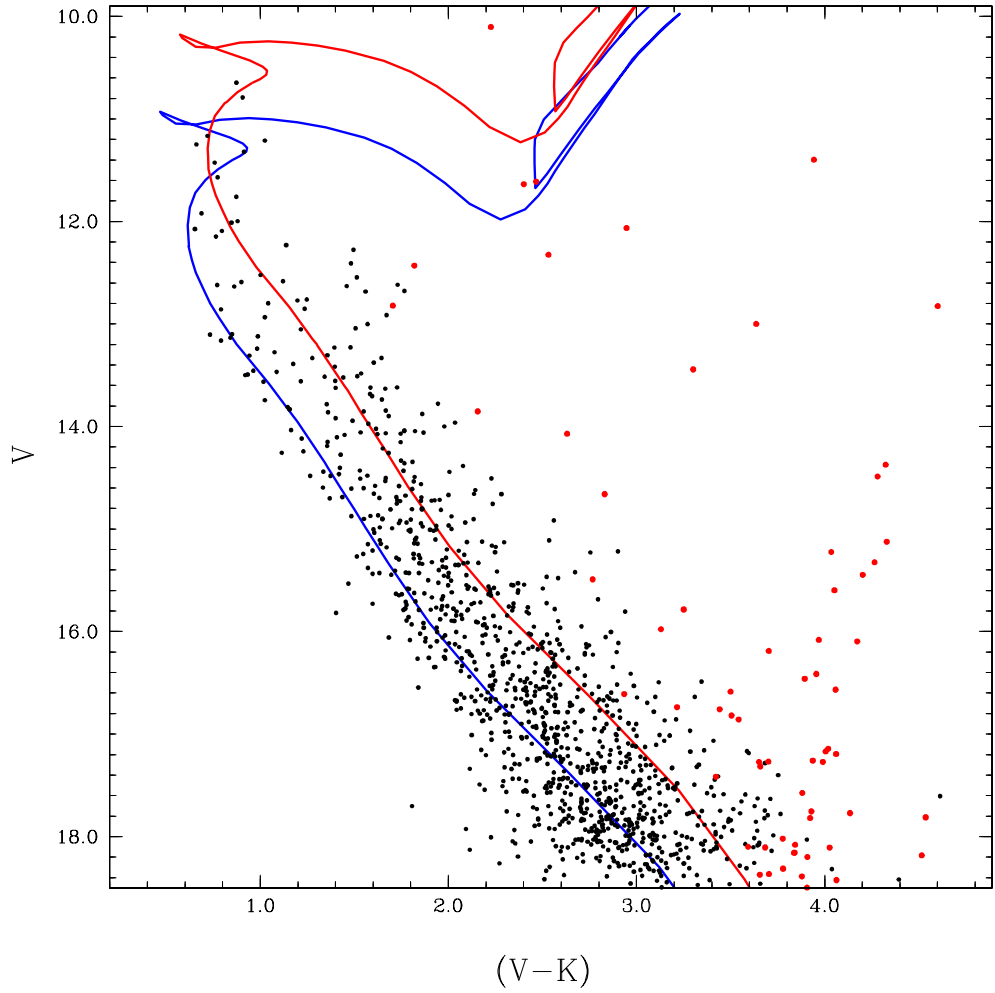}
\caption{The blue and red solid lines represent the boundary for binary stars. The blue line is also show the best fitted isochrone of the cluster whine the red line represents the red sequence of the binary stars.}
\label{s:fig08}
\end{figure}

The distance and age of any cluster is estimated through colour-magnitude diagrams (CMDs). These parameters of cluster had been already estimated by many authors \citep{jos12, gun12, gun13, jan14}. \cite{jan14} are computed the metallicity $Z=0.014{\pm}0.005$ for this cluster through Bayesian analysis, which shows close agreement with solar metallacity within error. Thus, we have adopted Marigo's theoretical isochrones of solar metallicity \citep{mar08} for further analysis. By taking the new reddening value ($0.12~mag$) and solar metallicity, we are fitted various isochrones of various age group nearby the literature ages of the cluster on ($V-K$ vs $V$) CMD. It is noticeable fact for any cluster that the scattering of stars is increases with the gap of effective wavelength of filters. We are familiar with the low detection rate of stars in $U$ and $B$ filters. In this background, $V$ band of the optical photometry and $K$ band of the 2MASS photometry were taken for constructing the CMD of the cluster.\\
Our best fitted isochrone solution provides the cluster log(age) and apparent distance modulus as $8.85{\pm}0.05$ and $10.75{\pm}0.10 mag$ respectively. The resultant distance modulus shows close agreement with the $(m-M)_V=10.98{\pm}0.24~mag$ \citep{jan14} while it is low compared to $11.15~mag$ \citep{jos12} and $11.33~mag$ \citep{kar05}. The best fitted isochrone has been shown by blue solid line in Figure~\ref{s:fig08}.
The red solid line in this Figure represents the red sequence which get by the adding $0.75~mag$ and $0.105~mag$ in  magnitude $(m-M)_V$ and colour ($V-K$) respectively. This shifting is taken account due to the  unresolved MS binaries. The heliocentric distance of the cluster is found to be $1.19{\pm}0.04~kpc$ through the following relation, $$d=10^{\left[(m-M)_V-R_V E(B-V) \right]/{5}+1}~kpc.$$\\
Whenever, we have change distance modulus 10.75 mag to 11.25 mag, we have noticed that stellar numbers between sequences is increases in $(B-V)/V$ and $(V-I)/V$ CMDs (see stellar distribution between dashed lines in Figure~\ref{s:fig10}) but deceases in  $(V-K)/V$ CMD (233 stars instead of 538 stars). This result leads an idea that the data incompleteness increases between $(B-V)/V$ and $(V-K)/V$ CMDs with the increasing distance modulus. The decreasing distance modulus shift the blue sequence in RCS region of $(B-V)/V$ CMD. Thus, the resultant apparent distance cluster is justified through this prescribed procedure.
\section{Blue Component Stars and Mass Function}\label{s:mf}
\begin{figure}
\includegraphics[width=13.5cm]{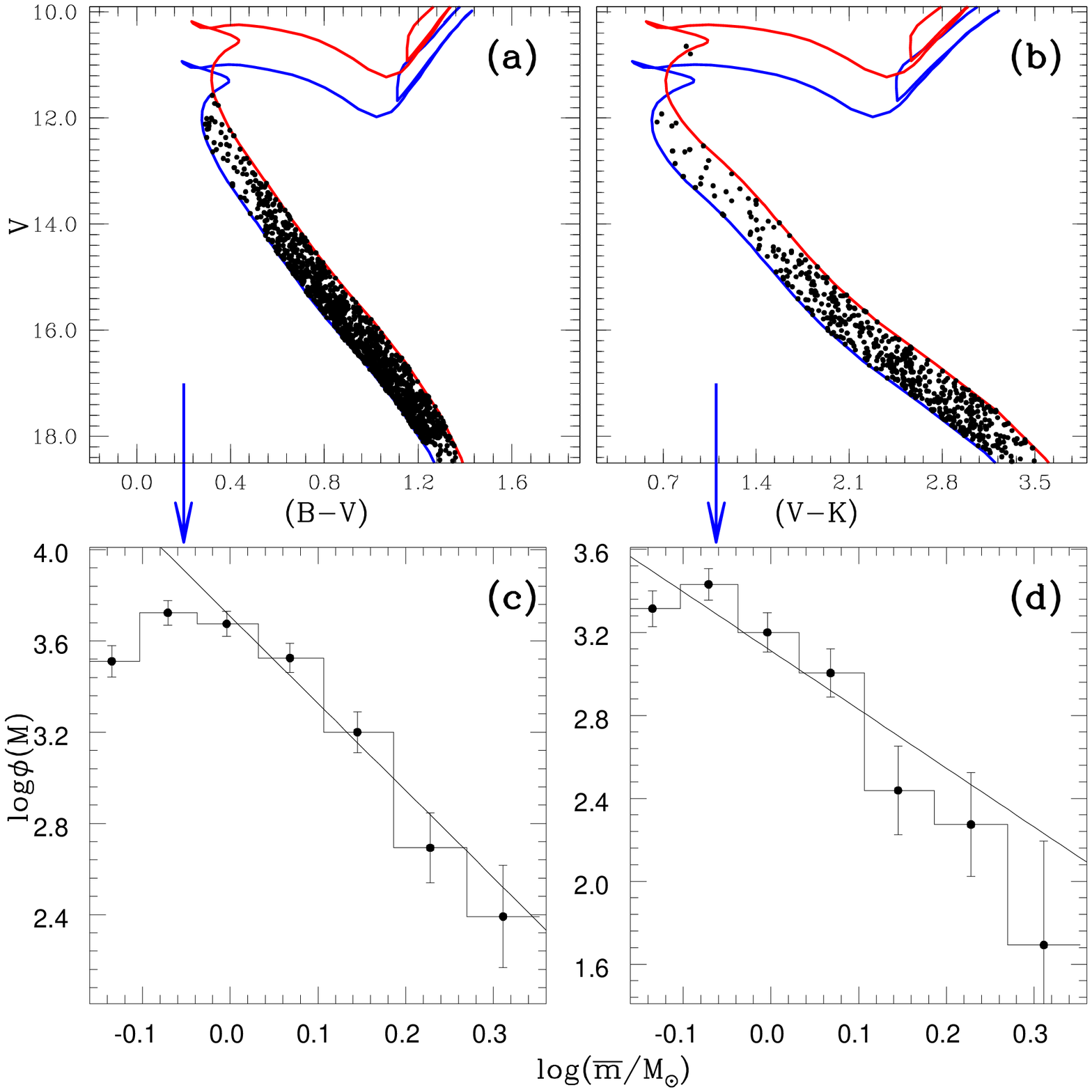}
\caption{The blue and red solid lines represent the boundary for binary stars in each CMD (a \& b). The MF slope values estimated through these CMDs have been shown in below of them (c \& d).}
\label{s:fig09}
\end{figure}

\begin{figure}
\includegraphics[width=13.5cm]{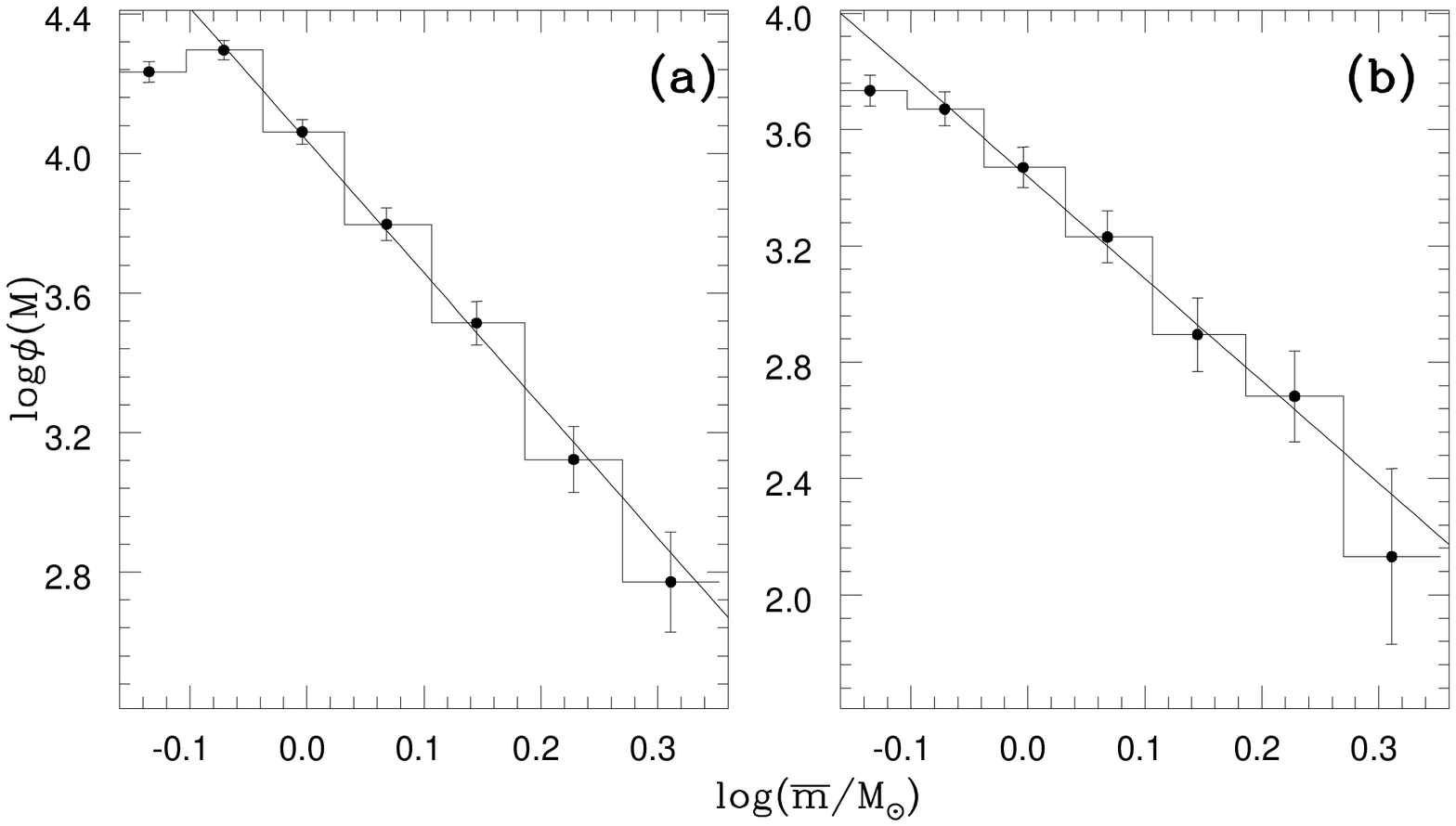}
\vspace{-3.7cm}
\caption{The MF slope values estimated through $(B-V)/V$ and $(V-K)/V$ CMDs.}
\label{s:fig10}
\end{figure}

\begin{figure}
\includegraphics[width=13.5cm]{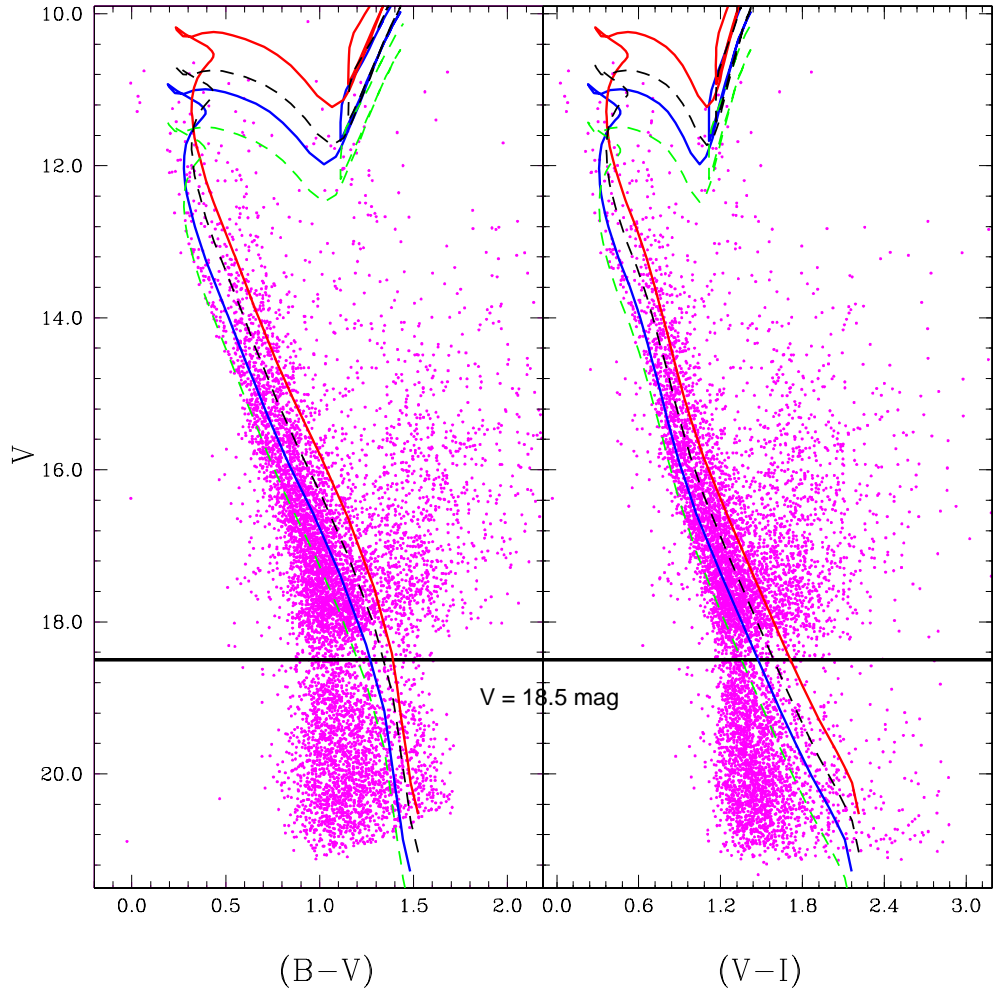}
\caption{The $(B-V)/V$ CMD and $(V-I)/V$ CMDs. The solid lines represent the boundary of blue and red sequence whenever the apparent distance modulus is 10.75 mag while dashed lines represent same for apparent distance modulus is 11.25 mag.}
\label{s:fig11}
\end{figure}

To understand the effect of incompleteness of data, we are used two type of CMDs for studying the mass function (MF) of the cluster. The MF is the relative number of stars per unit mass which is determined due to the absence a procedure of measurement of the initial mass function (IMF i.e. stellar mass distribution in per unit volume in a star formation event). The MF can be defined by a power law as $Nlog(M){\propto}M^{\Gamma}$ and its slope $\Gamma$ can be determined through  $$\Gamma=\frac{dlog~N~(log~m)}{d~log~m},$$ where $N~log(m)$ is the stellar number in per unit logarithm mass. Firstly, We are determined the MF slope of the cluster using the BCS fall between blue and red sequences. A total of 1318 stars in $(B-V)$ vs $V$ CMD and and 538 stars in $(V-K)$ vs $V$ CMD are found  between these sequences. The observed MF values have been given in Table~\ref{s:table1} and shown in Figure~\ref{s:fig09}.
\begin{table}
\caption{The MF of the cluster NGC 6866 derived from BCS between blue and red sequences in (A) $(B-V)$ vs $V$ CMD and (B) $(V-K)$ vs $V$ CMD (represented by suffix A and B respectively).}
\begin{center}
\begin{tabular}{@{}cccccc@{}}
\hline\hline
V range & Mass range & $\bar{m}$ & $log(\bar{m})$ & $N_{A}$ & $N_{B}$\\
(mag) & $M_{\odot}$  & $M_{\odot}$ & & &\\
\hline%
 11.5-12.5  & 2.235-1.854 & 2.045 &  0.311 &  20 &  04\\
 12.5-13.5  & 1.854-1.524 & 1.689 &  0.228 &  42 &  16\\
 13.5-14.5  & 1.524-1.267 & 1.395 &  0.145 & 127 &  22\\
 14.5-15.5  & 1.267-1.070 & 1.168 &  0.068 & 246 &  74\\
 15.5-16.5  & 1.070-0.912 & 0.991 & -0.004 & 328 & 110\\
 16.5-17.5  & 0.912-0.787 & 0.850 & -0.071 & 338 & 173\\ 
 17.5-18.5  & 0.787-0.677 & 0.732 & -0.135 & 212 & 135\\
\hline
\end{tabular}
\begin{tabular}{@{}ccccc@{}}
\hline
V range & $log(\Phi)_{A}$ & $e_{log(\Phi)_{A}}$ & $log(\Phi)_{B}$ & $e_{log(\Phi)_{B}}$\\
\hline%
 11.5-12.5  & 2.392 & 0.224 & 1.693 & 0.500 \\
 12.5-13.5  & 2.693 & 0.154 & 2.274 & 0.250 \\
 13.5-14.5  & 3.200 & 0.089 & 2.438 & 0.213 \\
 14.5-15.5  & 3.525 & 0.064 & 3.004 & 0.116 \\
 15.5-16.5  & 3.675 & 0.055 & 3.200 & 0.095 \\
 16.5-17.5  & 3.723 & 0.054 & 3.432 & 0.076 \\ 
 17.5-18.5  & 3.511 & 0.069 & 3.315 & 0.086 \\
\hline\hline
\end{tabular}
\end{center}
\label{s:table1}
\end{table}
The least stellar number in $(V-K)/V$ CMD indicates the data incompleteness. The $\Gamma$ values of BCS between sequences are computed to be $-3.81{\pm}0.51$ and $-2.83{\pm}0.69$ through $(B-V)/V$ and $(V-K)/V$ CMDs respectively. In present case, the incompleteness does not significantly altered the $\Gamma$ value of cluster though the mass function values are significantly decrease due to incompleteness of the data. The $\Gamma$ value through $(V-K)/V$ CMD is slightly increases which suggested that $\Gamma$ value increases due to the incompleteness of data.\\
The $\Gamma$ values are further estimated by all BCS through these CMDs. We are found 1111 BCS in  $(V-K)/V$ CMD which are low compare to 4088 BCS in $(B-V)/V$ CMD. The MF values through these members have been listed in Table~\ref{s:table2} and shown in Figure~\ref{s:fig10}.
\begin{table}
\caption{The MF of the cluster NGC 6866 derived from BCS found in (A) $(B-V)$ vs $V$ CMD and (B) $(V-K)$ vs $V$ CMD (represented by suffix C and D respectively).}
\begin{center}
\begin{tabular}{@{}ccccc@{}}
\hline\hline
V-mag & $N_{C}$ & $N_{D}$ & $log(\Phi)_{C}$ & $log(\Phi)_{D}$\\
\hline%
 11.5-12.5  &   48 &  11 & $2.772{\pm}0.114$ & $2.132{\pm}0.302$\\
 12.5-13.5  &  113 &  41 & $3.123{\pm}0.094$ & $2.683{\pm}0.156$\\
 13.5-14.5  &  262 &  63 & $3.514{\pm}0.062$ & $2.895{\pm}0.126$\\
 14.5-15.5  &  460 & 125 & $3.797{\pm}0.047$ & $3.231{\pm}0.089$\\
 15.5-16.5  &  800 & 205 & $4.062{\pm}0.035$ & $3.470{\pm}0.070$\\
 16.5-17.5  & 1265 & 300 & $4.296{\pm}0.028$ & $3.671{\pm}0.058$\\ 
 17.5-18.5  & 1121 & 355 & $4.234{\pm}0.030$ & $3.735{\pm}0.053$\\
\hline
\end{tabular}
\end{center}
\label{s:table2}
\end{table}
The ${\Gamma}$ values for all BCS are found to be $-3.80{\pm}0.11$ and $-3.51{\pm}0.19$ through $(B-V)/V$ and $(V-K)/V$ CMDs respectively. These results are show similar characteristic as mentioned above. To investigate why the mass function values in fainter end of both CMD have low value, we have saw the characteristic of the data of cluster. For this purpose, we have drawn  $(B-V)/V$ CMD and $(V-I)/V$ for cluster through all archive data of \cite{jan14} and these CMDs are shown in Figure~\ref{s:fig11}.\\
The very low stars are appeared between 18 to 19 mag of V-band as seen in Figure~\ref{s:fig11}. Therefore, the MF values found to be low compare to expected value. We have not found different ${\Gamma}$ for all BCS compare to that for BCS between sequences. Thus, we have considered the ${\Gamma}$ value for cluster $-3.80{\pm}0.11$ which is very high from to normal  ${\Gamma}$ value (-2.35) for clusters. We are also concluded that MF study is more reliable through BCS compare to the binary stars sequences as comes from the photometric criteria.    
\section{Nature of total-to-selective-extinction}\label{s:tcd2}
\begin{figure}
\includegraphics[width=13.5cm]{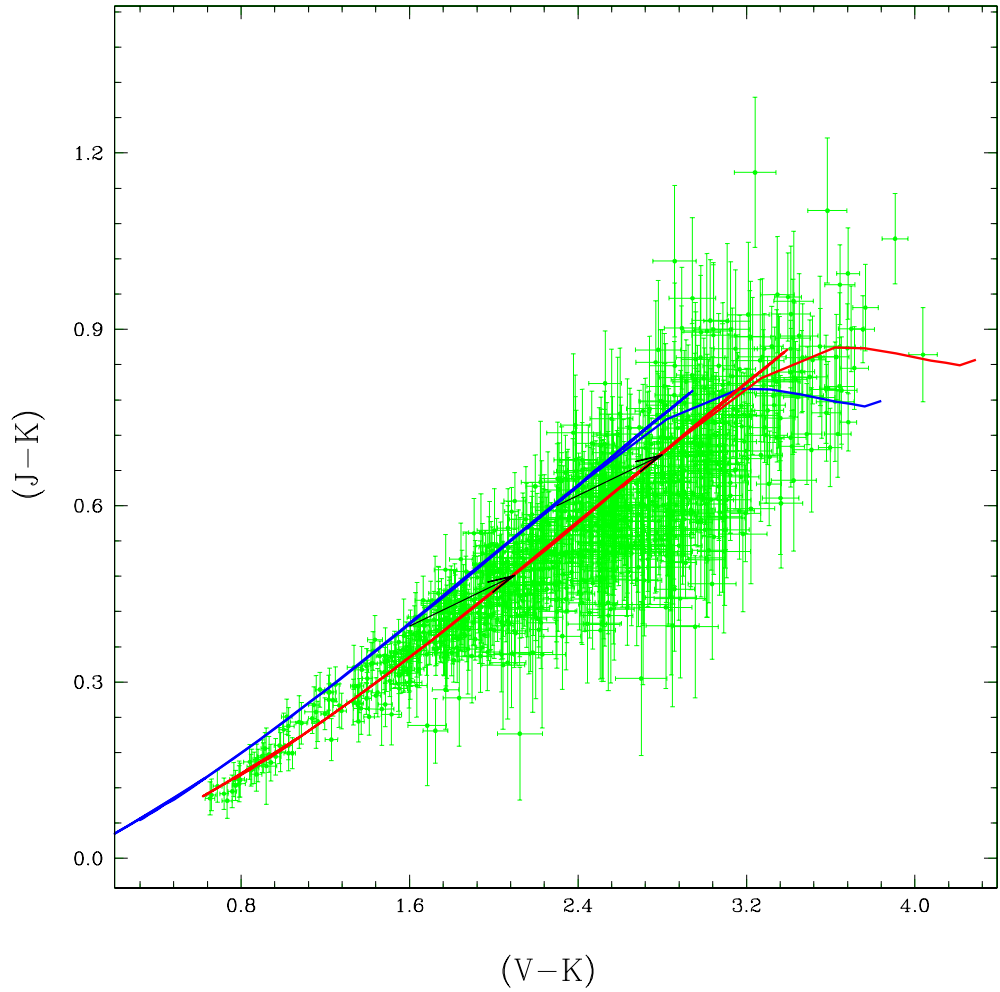}
\caption{The $(V-K)$ vs $(J-K)$ TCD for the cluster. The black solid arrows represent the normal reddening vector $\frac{E(J-K)}{E(V-K)}=0.173$. The horizontal and vertical green lines represent the errors in colours, (V-K) and (J-K).}
\label{s:fig12}
\end{figure}

We are estimated the IR interstellar extinction through the $(V-K)/(J-K)$ TCD as shown in Figure~\ref{s:fig12}.
In this figure, the blue solid line represents the solar metallicity isochrones of log(age)=8.85 while the red solid line is obtained by adding the 0.45 mag and 0.07 mag shift in $E(V-K)$ and $E(J-K)$, respectively. Thus, the colour-excess, $E(V-K)$ and $E(J-K)$ for the cluster are found to be 0.45 mag and 0.07 mag, respectively, which are give the value of normal reddening vector ($\frac{E(J-K)}{E(V-K)}$) as 0.16. This vector value is low than expected value i.e. 0.173. This low value is occurred due to the presence of interstellar clouds and dust in the direction of cluster field. 
The reddening value of cluster is found to be 0.16 mag through the \cite{mat90} relation, $\frac{E(V-K)}{E(B-V)}=2.75$. This resultant reddening value is slightly high compare to 0.12 mag as estimated through $(U-B)/(B-V)$ TCD.
These two different value of reddening through different CMD suggests that the total-to-selective-extinction value for the cluster is alter from the normal value.
\section{Conclusion}\label{s:con}
On the comparison of various photometric catalogue of this cluster, the photometric procedure of \citet{jos12} for estimation of the stellar magnitude seems to be less precise. Such least precision is occurred due to either the sky condition on the observational night or adopted standardization procedure. The photometric catalogue of \citet{jan14} is used to estimate the fundamental parameters of this cluster due to large observational field. Our deep inspection indicate that the radial over density of stars of cluster are occurred due to the BCS stars which is further supplemented by SNDs in various radial zones. The core-corona transition region is found as $7-10~arcmin$, which would be very effective to understand the evolutionary interaction between core and corona regions of the cluster. Though the BCS and RCS are neither distinguishable in RA-DEC plane nor distinct proper motions, whereas they have been separated on the basis of their location on $CMD$ plane. We have been also found that the ZAMS are not fitted on the RCS, whereas, RCS followed a linear trend. Our analysis shows that some RCS of core-region found to be Giant members of the cluster. The value of MF slope shows its dependency on the incompleteness of data. This manuscript is a mile-stone step to understand the exact members of the cluster through their known parametric values.
\section*{Acknowdege}\label{ss:ack}
This research has made use of the VizieR catalogue access tool, CDS, Strasbourg, France. The original description of the VizieR service was published in A{\&}AS 143, 23. GCJ is also thankful to APcyber Zone (Nanakmatta) for providing computer facilities. GCJ is also thankful to Shree Nilamber Joshi for providing the friendly environment, which becomes 
mile stone of my research work.
\bibliographystyle{model2-names}

\end{document}